\documentclass[11pt,twoside]{article}

\usepackage{asp2006}
\usepackage{epsf}
\usepackage{lscape}

\begin{document}
\title{Elemental Abundances in PG1159 Stars}   
\author{K. Werner,$^1$ T. Rauch,$^1$ E. Reiff,$^1$ and J. W. Kruk$^2$}  
\affil{$^1$Institute for Astronomy and Astrophysics, Kepler Center for
  Astro and Particle Physics, University of T\"ubingen, Sand~1, 72076 T\"ubingen, Germany\\
$^2$Department of Physics and Astronomy, Johns Hopkins University, Baltimore, MD 21218, U.S.A.
}  

\begin{abstract}
The hydrogen-deficiency in extremely hot post-AGB stars of spectral
class PG1159 is probably caused by a (very) late helium-shell flash or a
AGB final thermal pulse that consumes the hydrogen envelope, exposing
the usually-hidden intershell region. Thus, the photospheric elemental
abundances of these stars allow to draw conclusions about details of
nuclear burning and mixing processes in the precursor AGB stars. We
compare predicted elemental abundances to those determined by quantitative
spectral analyses performed with advanced non-LTE model atmospheres. A
good qualitative and quantitative agreement is found for many species
(He, C, N, O, Ne, F, Si, Ar) but discrepancies for others (P, S, Fe)
point at shortcomings in stellar evolution models for AGB stars. PG1159
stars appear to be the direct progeny of [WC] stars.

\end{abstract}

\begin{figure*}[bth]
\begin{center}
\epsfxsize=0.7\textwidth \epsffile{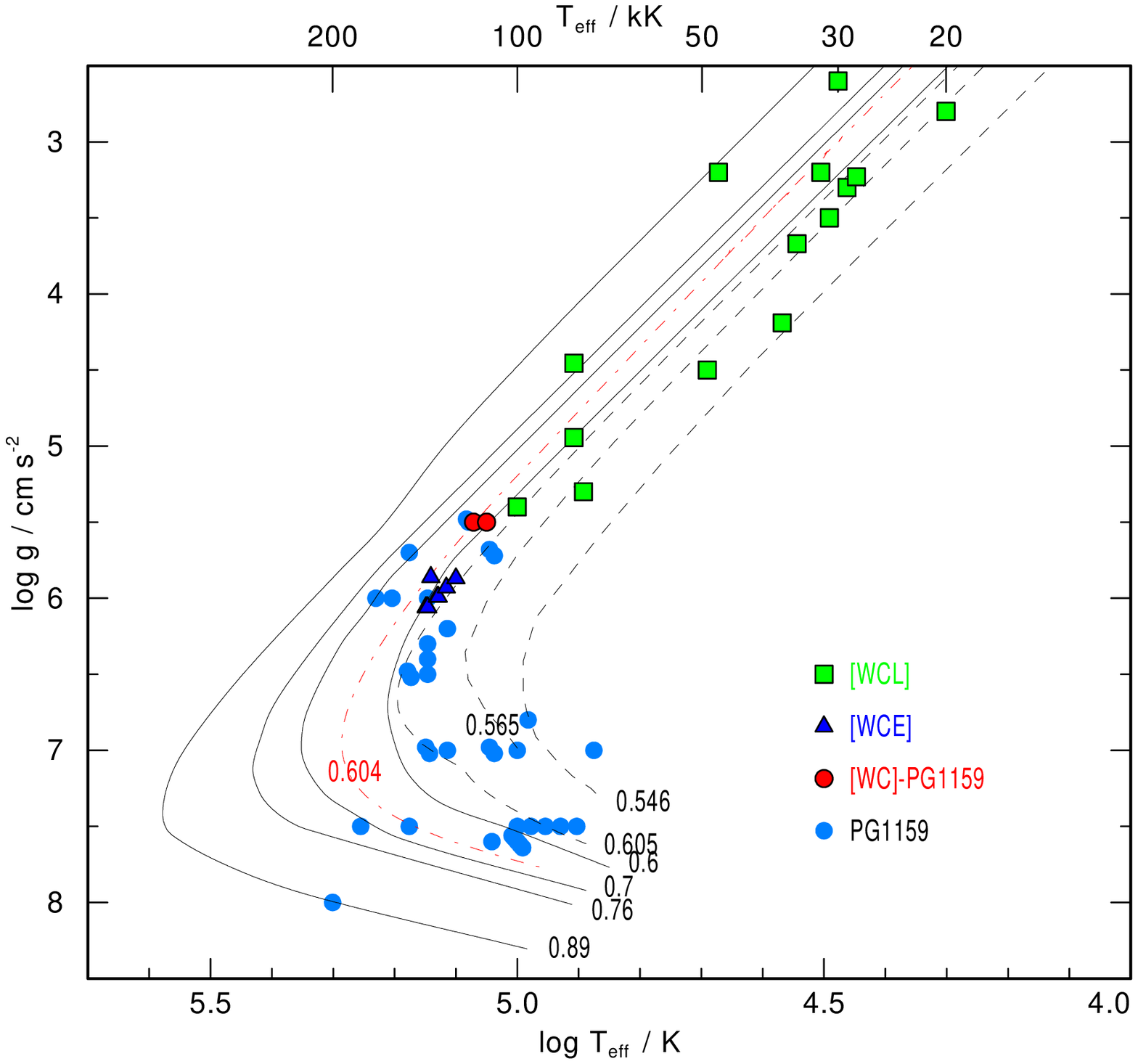}
\end{center}
\vspace{-0.5cm}
\caption{Hot hydrogen-deficient post-AGB stars in the
$g$--$T_{\rm eff}$--plane. We identify Wolf-Rayet central stars of early and
late type \citep[{[WCE], [WCL],} from][]{hamann:97}, PG1159 stars
\citep[from][]{werner:06} as well as two [WC]--PG1159 transition objects
(Abell~30 and 78). Evolutionary tracks are from \citet{schoenberner:83}
and \citet{bloecker:95b} (dashed lines), \citet{wood:86} and
\citet{2003IAUS..209..111H} (dot-dashed line; labels: mass in
M$_\odot$).  The latter 0.604\,M$_\odot$ track is the final CSPN track
following a VLTP evolution and therefore has a H-deficient
composition. However, the difference between the tracks is mainly due to
the different AGB progenitor evolution.}\label{fighrd}
\end{figure*}

\section{Main characteristics of PG1159 stars}

The PG1159 stars are a group of 40 extremely hot hydrogen-deficient
post-AGB stars.  Their effective temperatures ($T_{\rm eff}$) range
between 75\,000 and 200\,000~K. Many of them are still heating up along
the constant-luminosity part of their post-AGB evolutionary path in the
HR diagram ($L \approx 10^4$L$_\odot$) but most of them are already
fading along the hot end of the white dwarf cooling sequence (with $L$
{\raisebox{-0.4ex}{${\stackrel{>}{\scriptstyle \sim}} $}}
10\,L$_\odot$). Luminosities and masses are inferred from
spectroscopically determined $T_{\rm eff}$ and surface gravity ($\log
g$) by comparison with theoretical evolutionary tracks. The position of
analysed PG1159 stars in the ``observational HR diagram'', i.e., the
$T_{\rm eff}$--$g$ diagram, are displayed in Fig.\,\ref{fighrd}. The
high-luminosity stars have low $\log g$ ($\approx$\,5.5) while the
low-luminosity stars have a high surface gravity ($\approx$\,7.5) that
is typical for white dwarf (WD) stars. The derived mean mass is
0.57\,M$_\odot$, a value that is practically identical to the mean mass
of WDs \citep{be:07}. The PG1159 stars co-exist with hot central stars
of planetary nebulae and the hottest hydrogen-rich (DA) white dwarfs in
the same region of the HR diagram. About every other PG1159 star is
surrounded by an old, extended planetary nebula. For a recent review
with a detailed bibliography see \citet{werner:06}.

What is the characteristic feature that discerns PG1159 stars from
``usual'' hot central stars and hot WDs? Spectroscopically, it is the
lack of hydrogen Balmer lines, pointing at a H-deficient surface
chemistry. The proof of H-deficiency, however, is not easy: The stars
are very hot, H is strongly ionized and the lack of Balmer lines could
simply be an ionisation effect. In addition, every Balmer line is
blended by a Pickering line of ionized helium. Hence, only detailed
modeling of the spectra can give reliable results on the photospheric
composition. The high effective temperatures require non-LTE modeling of
the atmospheres. Such models for H-deficient compositions have only
become available in the early 1990s after new numerical techniques have
been developed and computers became capable enough.

\begin{figure*}[bth]
\vspace{-3.1cm}
\begin{center}
\epsfxsize=1.0\textwidth \epsffile{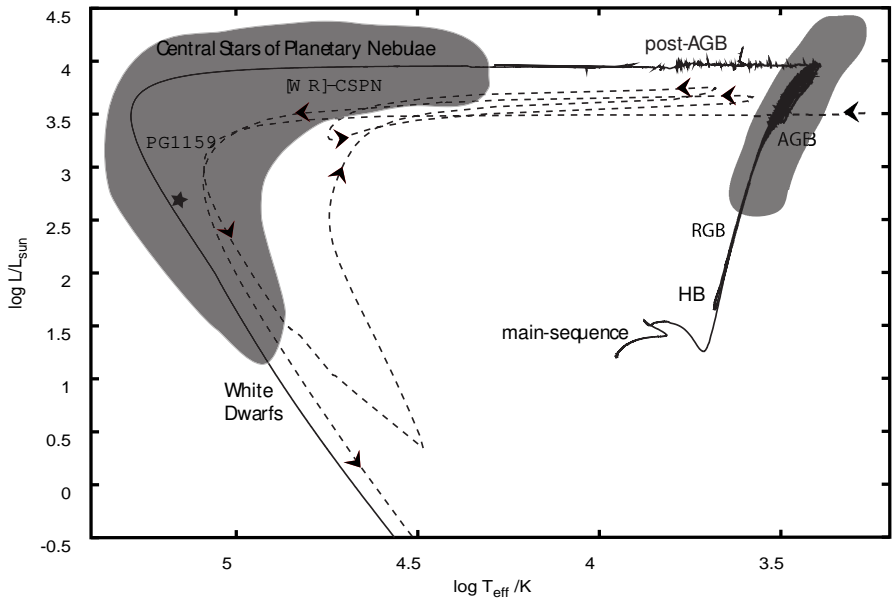}
\end{center}
\vspace{-6.9cm}
\caption{\label{fig:hrd} Complete stellar evolution track with an initial mass
of 2\,M$_\odot$ from the main sequence through the RGB phase, the HB to
the AGB phase, and finally through the post-AGB phase that includes the
central stars of planetary nebulae to the final WD stage. The solid line
represents the evolution of a H-normal post-AGB star. The dashed line
shows a born-again evolution of the same mass, triggered by a very late
thermal pulse, however, shifted by approximately $\Delta \log T_{\rm
eff} = -0.2$  and $\Delta \log~L/$L$_\odot = - 0.5$ for clarity. The
double-loop structure of the path is the consequence of a hydrogen-ingestion
flash. The
``$\star$'' symbol shows the position of PG1159$-$035
\citep[from][]{werner:06}.  
}
\end{figure*}                                                                                        

The first quantitative spectral analyses of optical spectra from PG1159
stars indeed confirmed their H-deficient nature \citep{werner:91}. It
could be shown that the main atmospheric constituents are C, He, and
O. The typical abundance pattern is C=0.50, He=0.35, O=0.15 (mass
fractions). It was speculated that these stars exhibit intershell matter
on their surface, however, the C and O abundances were much higher than
predicted from stellar evolution models. It was further speculated that
the H-deficiency is caused by a late He-shell flash, suffered by the
star during post-AGB evolution, laying bare the intershell layers. The
re-ignition of He-shell burning brings the star back onto the AGB,
giving rise to the designation ``born-again'' AGB star
\citep{iben:83a}. If this scenario is true, then the intershell
abundances in the models have to be brought into agreement with
observations. By introducing a more effective overshoot prescription for
the He-shell flash convection during thermal pulses on the AGB,
dredge-up of carbon and oxygen into the intershell can achieve this
agreement \citep{herwig:99c}. Another strong support for the born-again
scenario was the detection of neon lines in optical spectra of some
PG1159 stars \citep{werner:94}. The abundance analysis revealed Ne=0.02,
which is in good agreement with the Ne intershell abundance in the
improved stellar models.

If we do accept the hypothesis that PG1159 stars display former
intershell matter on their surface, then we can in turn use these stars
as a tool to investigate intershell abundances of other
elements. Therefore these stars offer the unique possibility to directly
see the outcome of nuclear reactions and mixing processes in the
intershell of AGB stars. Usually the intershell is kept hidden below a
thick H-rich stellar mantle and the only chance to obtain information
about intershell processes is the occurrence of the third
dredge-up. This indirect view onto intershell abundances makes the
interpretation of the nuclear and mixing processes very difficult,
because the abundances of the dredged-up elements may have been changed
by additional burning and mixing processes in the H-envelope (e.g.,
hot-bottom burning). In addition, stars with an initial mass below
1.5~M$_\odot$ do not experience a third dredge-up at all.

The central stars of planetary nebulae of spectral type [WC] are
believed to be immediate progenitors of PG1159 stars, representing the
evolutionary phase between the early post-AGB and PG1159 stages. This is
based on spectral analyses of [WC] stars which yield very similar
abundance results (see papers by Crowther, Todt, and Gr\"afener in these
proceedings).

\begin{figure*}
\begin{center}
\epsfxsize=1.0\textwidth \epsffile{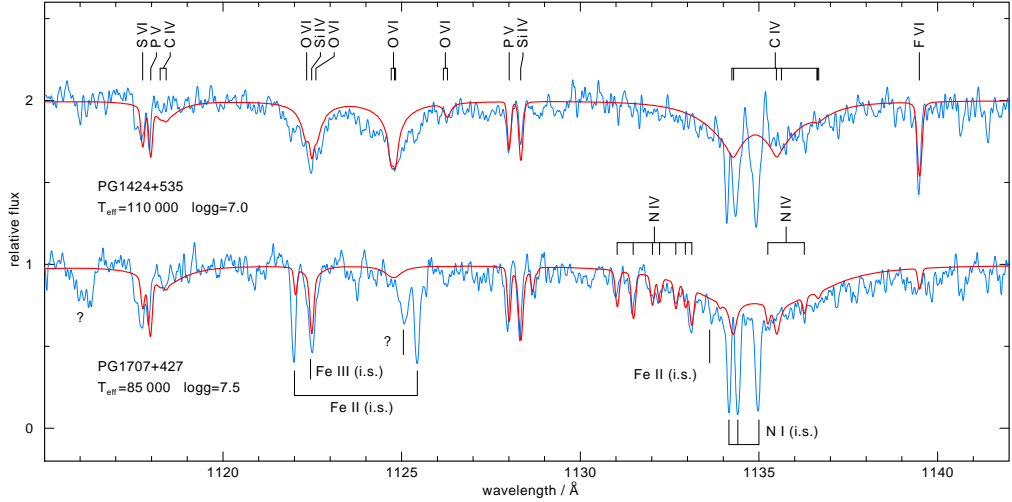}
\end{center}
\vspace{-0.5cm}
\caption{\label{fig:psifn} Detail from FUSE spectra of two relatively
  cool PG1159 stars (see labels). Note the
  following features. The
  \ion{F}{vi}~1139.5~\AA\ line which is the first detection of fluorine at all in
  a hot post-AGB star; the \ion{P}{v} resonance doublet at 1118.0 and
  1128.0~\AA, the first discovery of phosphorus in PG1159 stars; the
  \ion{N}{iv} multiplet at 1132~\AA. Also detected are lines from
  \ion{Si}{iv} and \ion{S}{vi}. The broader features stem from \ion{C}{iv}
  and \ion{O}{vi} \citep{reiff:07}.}
\end{figure*}                                                                                        

\section{Mass determination: spectroscopy and asteroseismology}

As mentioned, the mean spectroscopic mass of PG1159 stars is
0.57~M$_\odot$. This result was derived from the spectroscopic
temperature and gravity determination and comparison with modern
evolutionary tracks for H-deficient post-AGB stars \citep{mba:06}. This
is a systematic shift towards a lower mean mass compared to our
previously determined value (0.62~M$_\odot$) that was derived from
H-rich surface tracks \citep{werner:06}.

Some PG1159 stars are multi-periodic non-radial g-mode pulsators,
defining the group of GW~Vir stars (see, e.g., Quirion et al.\@ in these
proceedings). Stellar masses from asteroseismic modeling were derived
for five objects and we compare the results to the spectroscopic masses
in Table~\ref{tab:masses}.  Considering the uncertainties of both
methods we can claim consistent results, except for RX\,J2117.1+3412.

The result of an asteroseismic mass determination depends upon the
analysis method used. \citet{alt:08} discuss in detail the  results
derived from three methods, namely, the asymptotic and the average
period spacings, and detailed period fitting.  The spectroscopic mass
determination is most strongly affected by the uncertainty in
gravity. The accuracy is $\approx$0.3--0.5~dex, which propagates to an
error of the order 0.05--0.1~M$_\odot$, depending on the star's location
in the $g$--$T_{\rm eff}$--plane.  It is therefore fair to say that in
the current state the asteroseismology and spectroscopy results are of
comparable accuracy.

\begin{table}[t]
\begin{center}
\caption{
Spectroscopic vs.\@ asteroseismic masses. The
spectroscopic masses of RX\,J2117.1+3412 and the other four objects are
based on the temperature and gravity determinations by \citet{werner:96}
and \citet{dh:98}, respectively, and on the H-deficient evolutionary
tracks of \citet{mba:06}. The asteroseismic masses are from
\citet{alt:08}. The given ranges result from different pulsational
analysis methods.
\label{tab:masses}} 
\footnotesize
\begin{tabular}{lcc}
\noalign{\smallskip}
 \hline
\noalign{\smallskip}
Star            &M$_{\rm spec}/$M$_\odot$&M$_{\rm puls}$/M$_\odot$ \\
\hline
\noalign{\smallskip}
PG\,2131+066    & 0.55         & 0.58--0.63   \\ 
PG\,0122+200    & 0.53         & 0.56--0.62   \\
RX\,J2117.1+3412& 0.72         & 0.56--0.57   \\
PG\,1159$-$035  & 0.54         & 0.56--0.58   \\
PG\,1707+427    & 0.53         & 0.57--0.60   \\ 
\hline
\end{tabular}
\end{center}
\end{table}

Why should we care that the mass of a PG1159 star is
uncertain by 0.1~M$_\odot$? The answer is, that the predicted intershell
abundances strongly depend on the post-AGB remnant mass. That is
because the uncertainty in the respective main sequence progenitor mass
becomes large. According to the initial-final mass relation
\citep{weidi:00}, 0.60 and 0.68~M$_\odot$ remnants have evolved from
2.0 and 3.0~M$_\odot$ main sequence stars, respectively. E.g., depending on
metallicity, the predicted intershell fluorine abundance in these
stars can differ by an order of magnitude \citep{lugaro:04a}. Therefore,
if we want to use abundance patterns of PG1159 stars as a tool to
investigate AGB star nucleosynthesis, then we need a good mass
determination. In turn, if one believes that AGB star models describe
nucleosynthesis precisely enough, then the observed abundance pattern of
a PG1159 star can independently constrain its mass.

The recent discovery of a PG1159 star in a close binary system
(\citet{na:06}, and Schuh et al.\@ in these proceedings) might lead to a
dynamical mass determination, being an independent check for
the spectroscopic mass. Unfortunately, this star is a non-pulsator,
preventing an asteroseismic investigation.

\section{Three different late He-shell flash scenarios}

The course of events after the final He-shell flash is qualitatively
different depending on the moment when the flash starts. We speak about
a very late thermal pulse (VLTP) when it occurs in a WD, i.e., the star
had turned around the ``knee'' in the HR diagram and H-shell burning has
already stopped (Fig.\,\ref{fig:hrd}). The star expands and develops a
H-envelope convection zone that eventually reaches deep enough that
H-burning sets in (a so-called hydrogen-ingestion flash). Hence H is
destroyed and whatever H abundance remains, it will probably be shed off
from the star during the ``born-again'' AGB phase. A late thermal pulse
(LTP) denotes the occurrence of the final flash in a post-AGB star that
is still burning hydrogen, i.e., it is on the horizontal part of the
post-AGB track, before the ``knee''. In contrast to the VLTP case, the
bottom of the developing H-envelope convection zone does not reach deep
enough layers to burn H. The H-envelope (having a mass of about
$10^{-4}$~M$_\odot$) is mixed with a few times $10^{-3}$~M$_\odot$
intershell material, leading to a dilution of H down to about H=0.02,
which is below the spectroscopic detection limit. If the final flash
occurs immediately before the star departs from the AGB, then we talk
about an AFTP (AGB final thermal pulse). In contrast to an ordinary AGB
thermal pulse the H-envelope mass is particularly small. Like in the LTP
case, H is just diluted with intershell material and not burned. The
remaining H abundance is relatively high, well above the detection limit
(H {\raisebox{-0.4ex}{${\stackrel{>}{\scriptstyle \sim}} $}} 0.1).

\begin{figure}
\begin{center}
\epsfxsize=0.6\textwidth \epsffile{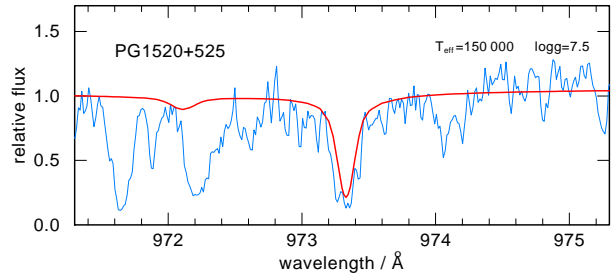}
\end{center}
\vspace{-0.5cm}
\caption{\label{fig:ne7} First identification of the
  \ion{Ne}{vii}~973.3~\AA\ line, shown here in the FUSE
  spectrum of the PG1159 star PG1520+525. This strong absorption feature
  is seen in the spectra many hot post-AGB stars, but remained
  unidentified for some years \citep{werner:04a}. }
\end{figure}                                                                                        

There are three objects, from which we believe to have witnessed a
(very) late thermal pulse during the last $\approx$\,100 years (details
on these stars are presented in several other papers of these
proceedings). FG~Sge suffered a late flash in 1894
\citep{gonzalez:98}. The star became rich in C and rare earth
elements. It most probably was hit by an LTP, not a VLTP, because it
turned H-deficient only recently (if at all, this is still under
debate). As of today, FG~Sge is located on or close to the AGB.
V605~Aql has experienced a VLTP in 1917 \citep{clayton:97}. Since then,
it has quickly evolved back towards the AGB, began to reheat and is now
in its second post-AGB phase. It has now an effective temperature of the
order 100\,000~K and is H-deficient.  Sakurai's  object (V4334~Sgr) also
experienced a VLTP, starting around 1993 \citep{duerbeck:96}. It quickly
evolved back to the AGB and became H-deficient. Recent observations
indicate that the reheating of the star has already begun, i.e., its second
departure from the AGB may be in progress.

The spectroscopic study of FG~Sge and Sakurai's object is particularly
interesting, because we can observe how the surface abundances change
with time. The stars are still cool, so that isotopic ratios can be
studied from molecule lines, and abundances of many
metals can be determined. The situation is less favorable with the hot
PG1159 stars: All elements are highly ionised and for many of them no
atomic data are available for quantitative analyses. On the other hand,
in the cool born-again stars the He-intershell material is once again
partially concealed.

\begin{figure}
\begin{center}
\epsfxsize=0.6\textwidth \epsffile{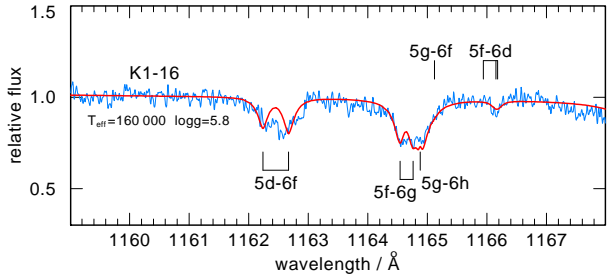}
\end{center}
\vspace{-0.5cm}
\caption{\label{fig:ne8} Discovery of \ion{Ne}{viii} lines in the FUSE
  spectrum of the PG1159-type central star of K\,1-16. This is the first
  detection of \ion{Ne}{viii} in any photospheric spectrum \citep{werner:07b}. Lines from this
  ion are only exhibited by the very hottest post-AGB stars ($T_{\rm eff}\geq$140\,000~K).}
\end{figure}                                                                                        

\section{Comparison of observed and predicted elemental abundances}

Abundance analyses of PG1159 stars are performed by detailed fits to
spectral line profiles. Because of the high $T_{\rm eff}$ all species
are highly ionized and, hence, most metals are only accessible by UV
spectroscopy. Optical spectra always exhibit lines from \ion{He}{ii} and
\ion{C}{iv}. Only the hottest PG1159 stars display additional lines of
N, O, and Ne (\ion{N}{v}, \ion{O}{vi}, \ion{Ne}{vii}). For all other
species we have utilized high-resolution UV spectra that were taken with
the \emph{Hubble Space Telescope} (HST) and the \emph{Far Ultraviolet
Spectroscopic Explorer} (FUSE). FUSE allowed observations in the
Lyman-UV range ($\approx$\,900--1200~\AA) that is not accessible with
HST. This turned out to be essential for most results reported here.

\begin{figure}
\begin{center}
\epsfxsize=0.6\textwidth \epsffile{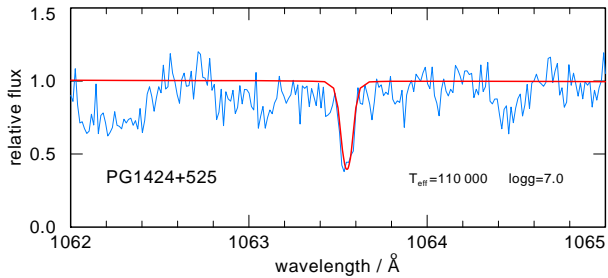}
\end{center}
\vspace{-0.5cm}
\caption{\label{fig:ar} Discovery of the \ion{Ar}{vii}~1063.55~\AA\ line in the FUSE
  spectrum of the PG1159 star PG1425+535. This is the first
  detection of \ion{Ar}{vii} in a photospheric spectrum and the only
  detectable argon line in any wavelength region of hot post-AGB stars \citep{werner:07a}.}
\end{figure}

\begin{figure}
\begin{center}
\epsfxsize=0.6\textwidth \epsffile{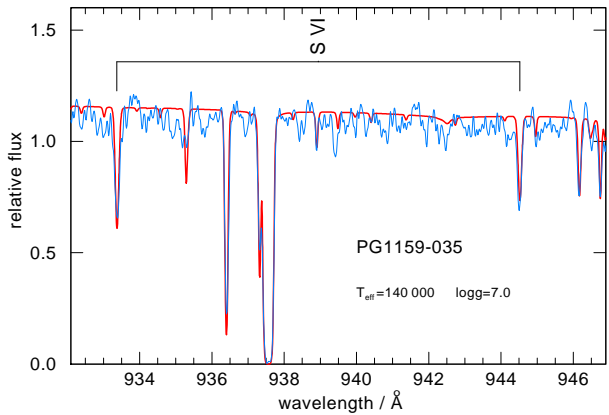}
\end{center}
\vspace{-0.5cm}
\caption{\label{fig:s6} FUSE enabled the first abundance determination of sulfur in
  PG1159 stars. Shown here is the \ion{S}{vi} resonance doublet in the
  prototype of the PG1159 spectral class, PG1159$-$035 \citep{jahn:07}.}
\end{figure}

\begin{figure}
\begin{center}
\epsfxsize=0.6\textwidth \epsffile{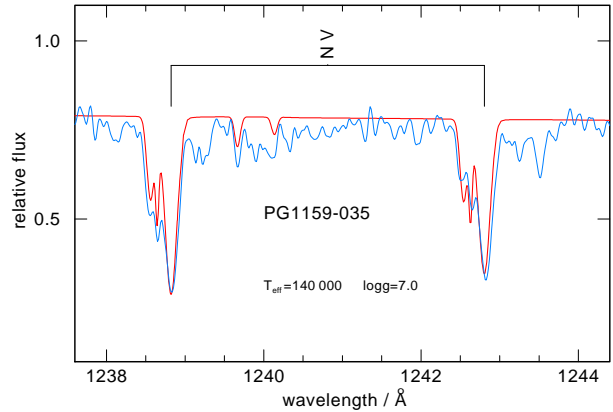}
\end{center}
\vspace{-0.5cm}
\caption{\label{fig:n5} The high spectral resolution capability of STIS
  aboard HST allows to distinguish the photospheric \ion{N}{v} resonance
  doublet from the weak blueshifted ISM components. This enabled the first reliable
  nitrogen abundance determination in the prototype PG1159$-$035 \citep{jahn:07}.}
\end{figure}

A number of chemical elements could be identified for the very first
time (F, P, S, Ar). In addition, very high ionisation stages of several
elements, which were never seen before in stellar photospheric spectra,
could be identified in the UV spectra for the very first time
(e.g. \ion{Si}{v}, \ion{Si}{vi}, \ion{Ne}{viii}). To illustrate this, we
display in Figs.~\ref{fig:psifn}--\ref{fig:n5} details of FUSE and HST
spectra of PG1159 stars.  We discuss the spectroscopic results comparing
with [WC]s and model predictions.

\emph{Hydrogen --} Four PG1159 stars show residual H with an abundance
of 0.17 (so-called hybrid-PG1159 stars). These objects are the outcome
of an AFTP. All other PG1159 stars have H
{\raisebox{-0.4ex}{${\stackrel{<}{\scriptstyle \sim}} $}} 0.1 and,
hence, should be LTP or VLTP objects. H was also found in [WCL] stars
(H=0.01--0.1).

\emph{Helium, carbon, oxygen --} These are the main constituents of
PG1159 atmospheres. A large variety of relative He/C/O abundances is
observed. The approximate abundance ranges are: He=0.30--0.85,
C=0.13--0.60, O=0.02--0.20. The spread of abundances might be explained
by different numbers of thermal pulses during the AGB phase, except for
the most He-rich stars. They might belong to a different post-AGB
sequence involving the so-called O(He) stars (see Rauch et al.\@ in
these proceedings). The He/C/O abundances in PG1159 stars are
consistent with results found for [WC] stars.

\emph{Nitrogen --} N is a key element that allows to decide if the star
is the product of a VLTP or a LTP. Models predict that N is diluted
during an LTP so that in the end N=0.1\%. This low N abundance is
undetectable in the optical and only detectable in extremely good UV
spectra. In contrast, a VLTP produces N (because of H-ingestion
and burning) to an amount of 1\% to maybe a few percent. N abundances of
the order 1\% are found in some PG1159 stars, while in others it is
definitely much lower. This picture is similar to the [WC] stars.

\emph{Neon --} Ne is made from $^{14}$N that was produced by CNO
burning. In the He-burning region, $\alpha$-captures transform
$^{14}$N to $^{22}$Ne. Evolution models predict Ne=0.02 in the
intershell. A small spread is expected as a consequence of different
initial stellar masses. Ne=0.02 was first found from optical analyses of a
few stars and, later, in a larger sample observed with FUSE
\citep{werner:04a}. The Ne abundance in [WC] stars is very similar (0.02--0.04).

\emph{Fluorine --} F was for the first time discovered by
\citet{werner:05} in hot post-AGB stars; in PG1159 stars as well as
hydrogen-normal central stars. A strong absorption line in FUSE spectra
located at 1139.5~\AA\ remained unidentified until we found that it
stems from \ion{F}{vi}. The abundances derived for PG1159 stars show a
large spread, ranging from solar to up to 250 times solar. This was
surprising at the outset because $^{19}$F, the only stable F isotope, is
very fragile and easily destroyed by H and He. A comparison with AGB
star models of \citet{lugaro:04a}, however, shows that such high F
abundances in the intershell can indeed be accumulated by the reaction
$^{14}$N($\alpha$,$\gamma$)$^{18}$F($\beta^+$)$^{18}$O(p,$\alpha$)$^{15}$N($\alpha$,$\gamma$)$^{19}$F,
the amount depends on the stellar mass. We find a good agreement between
observation and theory. Our results also suggest, however, that the F
overabundances found in AGB stars \citep{jorissen:92} can only be
understood if the dredge-up in AGB stars is much more efficient
than hitherto thought.

\emph{Silicon --} Si is expected to remain almost
unchanged, in agreement with PG1159 stars for which we could
determine the Si abundance. The same holds for some [WC] stars, but in
other cases overabundances were found (8--45 times solar).

\emph{Phosphorus --} Systematic predictions from evolutionary model
grids are not available; however, the few computed models show P
overabundances in the range 4--25 times solar (Lugaro priv.\@ comm.). This
is at odds with our spectroscopic measurements for two PG1159 stars,
that reveal a solar P abundance.

\emph{Sulfur --} Again, model predictions are uncertain at the
moment. Current models show a slight (0.6 solar) underabundance. In
strong contrast, we find a large spread of S abundances in PG1159 stars,
ranging from solar down to 0.01 solar.

\emph{Argon --} Ar was identified recently for the
first time in hot post-AGB stars and white dwarfs \citep{werner:07a}. Among them is one
PG1159 star for which a solar abundance has been determined (Fig.\,\ref{fig:ar}). This is
in agreement with AGB star models which predict that the Ar abundance
remains almost unchanged.

\begin{figure}
\begin{center}
\epsfxsize=0.6\textwidth \epsffile{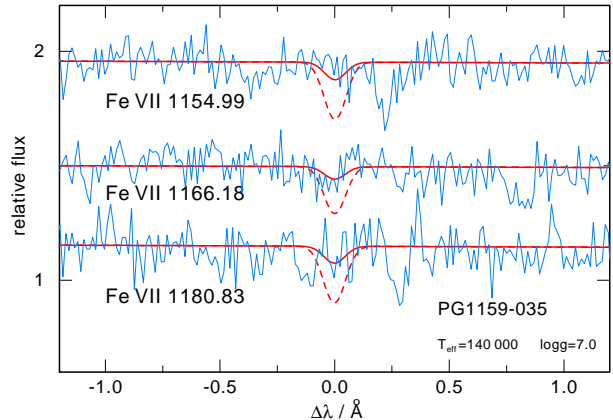}
\end{center}
\vspace{-0.5cm}
\caption{\label{fig:fe} From the nondetection of \ion{Fe}{vii} lines in
  PG1159$-$035 a Fe deficiency is found. Here we
  compare model spectra for three lines, each computed with solar and 0.1 solar Fe
  abundance, to the observation \citep{jahn:07}.}
\end{figure}

\emph{Iron and Nickel --} \ion{Fe}{vii} lines are expected to be the
strongest iron features in PG1159 stars. They are located in the UV
range. One of the most surprising results is the non-detection of these
lines in three examined PG1159 stars (K1-16, NGC\,7094, PG1159$-$035;
see, e.g., Fig.\,\ref{fig:fe}). The derived upper abundance limits
\citep[e.g.][]{werner:03,jahn:07} indicate that iron is depleted by
about 0.7--2 dex, depending on the particular object. Iron depletions
were also found for the PG1159-[WC] transition object Abell~78 as well
as for several PG1159 progenitors, the [WC] stars. It must be stressed
that in no single case a solar (or almost solar) Fe abundance was
found. Such high Fe depletions are not in agreement with current AGB
models. Destruction of $^{56}$Fe by neutron captures is taking place in
the AGB star intershell as a starting point of the s-process; however,
the resulting depletion of Fe in the intershell is predicted to be small
(about 0.2~dex). It could be that additional Fe depletion can occur
during the late thermal pulse. In any case, we would expect a
simultaneous enrichment of nickel, but up to now we were unable to
detect Ni in PG1159 stars at all (see Reiff et al.\@ in these
proceedings).

A particularly mysterious problem is the Fe-deficiency in
the hybrid-PG1159 central star NGC\,7094 (Ziegler et al., these
proceedings). We recall that the hybrid-PG1159 stars are the outcome of
an AFTP with residual hydrogen (H$ = 0.17 $) from the envelope that was mixed with intershell
matter. One would expect to see at least the iron that was contained in the
convective H-envelope, hence Fe/H $\approx$\,solar. This is not the case:
the upper limit for Fe/H ratio is much smaller.

\emph{Trans-iron elements --} The discovery of s-process elements would
be highly desirable. However, this is  impossible due to the lack of
atomic data. From the ionization potentials we expect that these
elements are highly ionised like iron, i.e., the dominant ionization
stages are \ion{}{vi}~--\ion{}{ix}. To our best knowledge, there are no
laboratory measurements of so highly ionised species that would allow us
to search for atomic lines in the spectra. Such measurements would be
crucial to continue the elemental abundance determination beyond the
current state.

\section{The enigmatic case of H1504+65}

H1504+65 is the hottest PG1159 star ($T_{\rm eff}=200\,000$~K, $\log
g=8$), and it is H- \emph{and} He-deficient \citep{werner:04b}. The
atmosphere is dominated by C and O (C=0.48, O=0.48, Ne=0.02,
Mg=0.02). It is the most massive PG1159 star
(0.74--0.97\,M$_\odot$). However, its evolutionary history is completely
unknown, and the model tracks used to derive the mass might be
inappropriate. It was speculated that H1504+65 is a bare CO or even a
ONeMg WD, so that the progenitor could have been a super-AGB star. In
any case, the surface chemistry is a real challenge for evolution
theory. Until now, H1504+65 appeared as a ``singularity'', with no
potential progenitor or progeny candidates. The recently discovered cool
DQ WDs with almost pure C atmospheres, however, could represent such a
progeny (Dufour et al., these proceedings).

\section{Conclusions}

It has been realized that PG1159 stars exhibit intershell matter on
their surface, which has probably been laid bare by a late final thermal
pulse. This provides the unique opportunity to study directly the result
of nucleosynthesis and mixing processes in AGB stars. Abundance
determinations in PG1159 stars are in agreement with intershell
abundances predicted by AGB star models for many elements (He, C, N, O,
Ne, F, Si, Ar). For other elements, however, disagreement is found (Fe,
P, S) that points at possible weaknesses in the evolutionary
models. Generally, the abundance patterns clearly support the idea that
[WC] stars are direct progenitors of PG1159 stars.

\acknowledgements TR is supported by the \emph{German Astrophysical
Virtual Observatory} project of the German Federal Ministry of Education
and Research (grant 05\,AC6VTB). ER is supported by DFG grant
We\,1312/30-1. JWK is supported by the FUSE project, funded by NASA
contract NAS5-32985.

\end{document}